\documentclass[prx,aps,letterpaper,twocolumn,allowfontchageintitle=true, accepted=2024-05-15]{quantumarticle}
\pdfoutput=1

\usepackage[table]{xcolor}
\usepackage{amsmath,amsfonts,amssymb,graphics,graphicx,epsfig,color,times,bbm}
\usepackage{amsthm}
\usepackage{mathtools}
\usepackage{psfrag}
\usepackage{braket}
\usepackage{hyperref}
\usepackage{wasysym}
\usepackage{bbm}
\usepackage{appendix}
\usepackage{hyperref}
\usepackage{lipsum}
\usepackage{booktabs}
\usepackage{natbib}
\usepackage{graphicx}
\usepackage{tabulary}

\graphicspath{{figures/}}

\newcommand{\id}{\mathbbm{1}}

\begin{document}

\title{\centering \color{quantumviolet} Unifying flavors of fault tolerance with the ZX calculus \newpage }

\author{\centering Hector Bombin}
\author{Daniel Litinski} 
\author{Naomi Nickerson}
\author{Fernando Pastawski}
\author{Sam Roberts \vspace{-2ex}}
\affiliation{PsiQuantum, Palo Alto\vspace{-1ex}}
\date{\vspace{-4ex}}

\begin{abstract}
    There are several models of quantum computation which exhibit shared fundamental fault-tolerance properties.
    This article makes commonalities explicit by presenting these different models in a unifying framework based on the ZX calculus.
    We focus on models of topological fault tolerance~--~specifically surface codes~--~including circuit-based, measurement-based and fusion-based quantum computation, as well as the recently introduced model of Floquet codes.
    We find that all of these models can be viewed as different flavors of the same underlying stabilizer fault-tolerance structure, and sustain this through a set of local equivalence transformations which allow mapping between flavors.
    We anticipate that this unifying perspective will pave the way to transferring progress among the different views of stabilizer fault-tolerance and help researchers familiar with one model easily understand others.
\end{abstract}

\maketitle

\begin{figure*}
	\centering
	\includegraphics[width=0.95\linewidth]{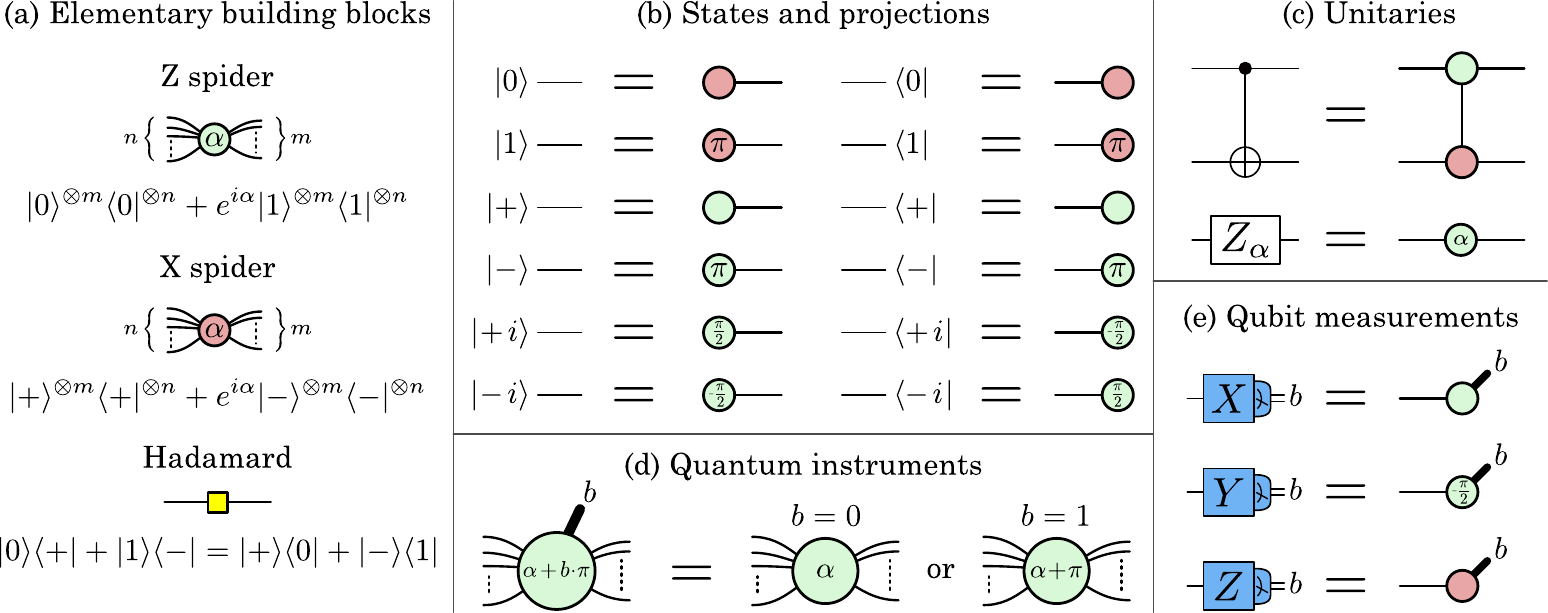}
	\caption{(a) $Z$ spiders, $X$ spiders and Hadamards are the elementary building blocks of ZX diagrams. (b) Spider nodes with one leg can be used to describe Pauli eigenstates and their projections. (c) Unitary gates can also be expressed as ZX diagrams, as shown for the CNOT gate and a single-qubit $Z$ rotation. (d) Classical outputs can be added to describe quantum instruments. 
	The classical bit $b$ determines whether the phase of the spider is $\alpha$ or $\alpha + \pi$. (e) Such ZX instruments can be used to describe Pauli measurements.}
\label{fig:zxintro1}
\end{figure*}

\section{Introduction}
The introduction of Kitaev's toric code in 1996 \cite{kitaev2003} has spurred a host of research in fault-tolerant quantum computation and topological condensed-matter physics.
Whereas a quantum error-correcting code describes how information is protected at a given instant (snapshot), fault-tolerant computing describes an interactive process through which classical control can repeatedly operate on  encoded quantum information without ever leaving relevant observables  unprotected.
In this article, we compare the description of quantum memory, the simplest fault-tolerant primitive, as described in the following four models:

\textbf{1. Circuit-based quantum computation (\textbf{CBQC}).} In this model~\cite{dennis2002}, the quantum computer is an array of physical qubits, often arranged on a 2D grid. It is possible to perform single-qubit gates and measurements, as well as entangling two-qubit gates, usually between nearest neighbors in the grid. A quantum memory is implemented by using these operations to repeatedly measure the surface-code stabilizer operators. The measurement outcomes are used to detect and correct errors corrupting the physical qubits, as well as errors in the measurement outcomes. In the absence of noise and errors, all stabilizer measurement outcomes involved in fault tolerance have predetermined outcomes, i.e. there is no intrinsic randomness in the outcomes.

\textbf{2. Measurement-based quantum computation (MBQC).} In this model~\cite{raussendorf2003, briegel2009,raussendorf2006}, a quantum computation consists of two steps. First, a large initial entangled state is prepared. Roughly speaking, the number of physical qubits in this model corresponds to a product of the number of physical qubits in the circuit model multiplied by the number of circuit time steps for which they are present. The initial entangled state, usually referred to as the cluster state, can be prepared from a product state by a constant-depth Clifford circuit which is, to a large extent, independent of the target computation. In the second step, qubits are consumed by single-qubit measurements as the computation progresses. 
Measurement outcomes in fault-tolerant MBQC are used to diagnose errors in both preparation and measurement. 
A feature which distinguishes MBQC from CBQC is that individual measurement outcomes are intrinsically random, even in a noiseless setting.

\textbf{3. Fusion-based quantum computation (\textbf{FBQC})}~\cite{bartolucci2021}. Here, a quantum computer is thought of as a collection of resource-state generators that repeatedly produce copies of small few-qubit entangled states (\textit{resource states}). 
The size of these resource states is constant and does not depend on the quantum algorithm or the code distance. 
In contrast, the total number of resource states  will scale with the size of the computation.
Complementary to resource-state generation is resource-state consumption, referred to as fusion, which consists of entangling two-qubit (or multi-qubit) measurements between two (or more) such resource states. This can be used to successively teleport the encoded logical state into fresh resource states and advance the computation.
As with teleportation, many of the measurement outcomes are expected to be individually random.
Characteristic to this approach is that qubits are dynamically created and destroyed throughout the computation by resource state generation and by fusion measurements respectively.
This is naturally suited to photonic qubits which are destroyed after measurement.

\textbf{4. Floquet-based quantum computation (FloBQC)}. This is a relatively recent development also referred to as pair-measurement codes~\cite{hastings_2021, paetznick2022,haah_2022, Gidney2022,Gidney2022b,Kesselring2022}.
As in CBQC, the base model assumes a fixed set of physical qubits which evolve in time with the aid of projective parity measurements on qubit pairs.
However, as in FBQC, these schemes yield measurement outcomes which are random when viewed independently and only two-qubit measurements are present.
This model of computation is motivated by the elusive Majorana qubits \cite{Aguado2020}, on which
researchers expect joint parity measurement, rather than entangling unitary gates, to be natural operations.

In this manuscript we clarify how the stated collection of protocols are part of the same happy family of topological stabilizer fault tolerance, even though the four models of computation feature completely different physical operations. 
A key aspect of our viewpoint is the importance of thinking of fault tolerance holistically in space-time, rather than simply as operations on a code. 
This viewpoint is prominently advocated for in, e.g., Refs.~\cite{Bombin2021, gottesman2022opportunities}. 
In this view, checks are characterized by the classical outcomes they are derived from and by the errors which can flip them. 
After providing a self-contained introduction to ZX diagrams with some small additions for the purposes of fault tolerance, we apply these tools to the four different models of computation.

\section{ZX tensor network diagrams}\label{sec:ZX}
In order to identify the commonalities between these fault-tolerant protocols, it is necessary to use a shared language to describe them. 
{\it Tensor networks}  provide a convenient common language for this.
As is common in tensor network diagrams, edges between tensor nodes represent  contraction.
The specific notation of ZX diagrams \cite{Coecke2011, Backens2014, Coecke2017,  Beaudrap2020} provides a concise way to graphically represent not just the tensor signature, but also the tensor {\it content}.
While ZX diagrams have an extremely limited number of elementary building blocks, these can be combined (contracted) to represent arbitrary composite qubit tensors \cite{Hadzihasanovic2018, Jeandel2018}.
As we will see, for many relevant tensors on qubits these representations can be remarkably succinct, a point which has already been made in the context of surface code lattice surgery \cite{Beaudrap2020}.
Moreover,  small diagram equalities can be used as graphical rewriting rules to transform ZX diagrams into equivalent representations. 
This sections provides a brief self-contained summary of the ZX diagram features used, a topic for which there already exists good comprehensive documentation~\cite{Coecke2017, Wetering2020}. 

\textbf{Elementary ZX tensor nodes.} 
The main protagonists (elementary tensors or linear maps) of ZX tensor network diagrams are $X$-spiders (usually colored red or sometimes gray) and $Z$-spiders (usually colored green or sometimes white), and are shown in Fig.~\ref{fig:zxintro1}a.
These are both denoted by circular nodes and may have an arbitrary number of legs corresponding to qubit ports attached to them. 
Spider nodes are parameterized by a phase $\alpha$ which is assumed to be zero unless explicitly labeled on the node.
A third commonly used tensor is the Hadamard unitary tensor $H = |0\rangle \langle +| + |1\rangle \langle -|$, which has exactly two ports and is represented by a square node (usually colored yellow) with one port on each side.

\begin{figure*}
	\centering
	\includegraphics[width=\linewidth]{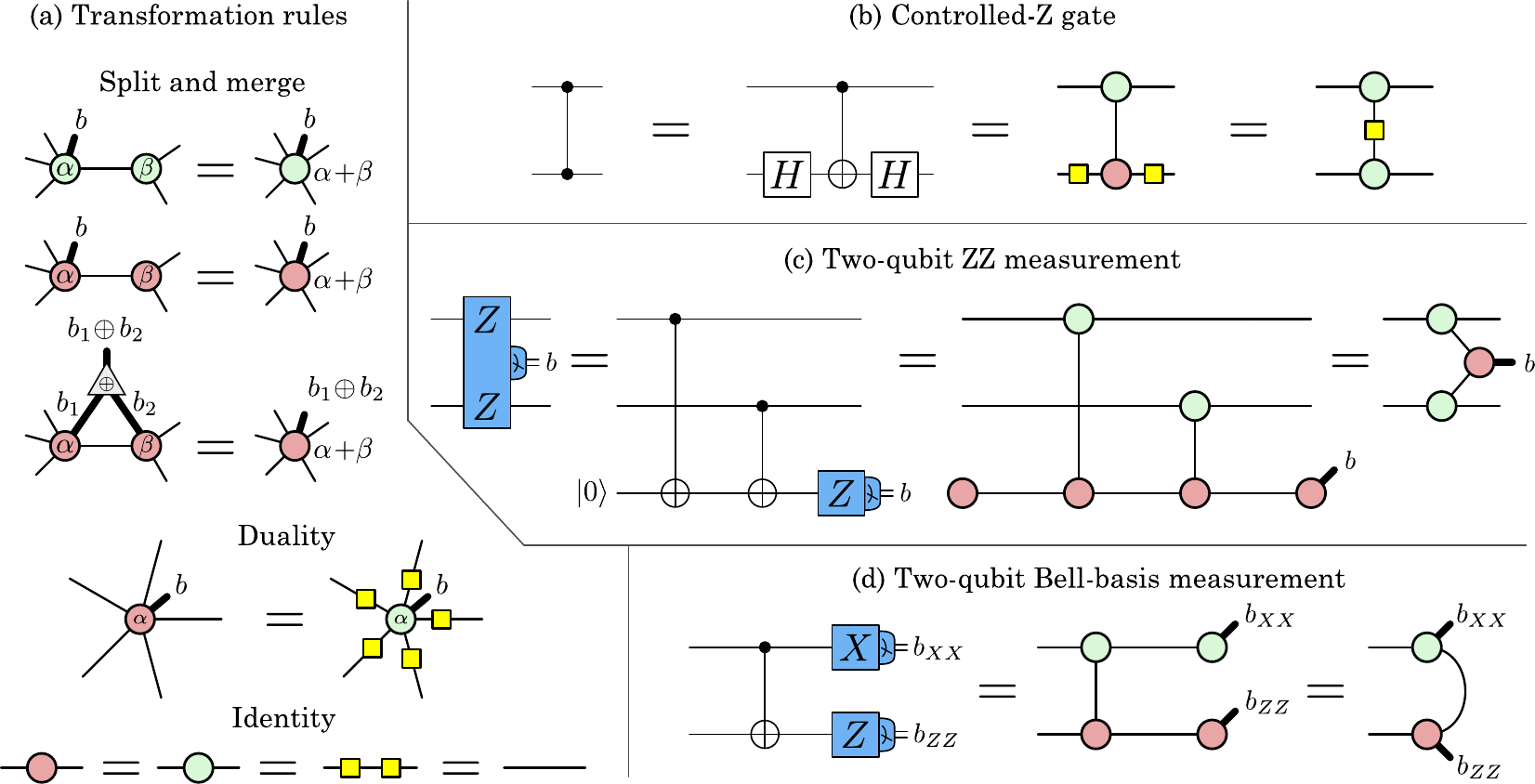}
	\caption{(a) The most commonly used ZX-diagrammatic equations. These identities are also valid without classical outputs, i.e., by removing all outputs indicated by thick black lines. The transformation rules can be used to simplify (b) the \texttt{CZ} gate, (c) the circuit of a non-destructive $ZZ$ measurement and (d) the circuit of a destructive two-qubit Bell measurement.}
\label{fig:zxintro2}
\end{figure*}

\textbf{States, projections and unitary gates.} Figure \ref{fig:zxintro1}b enumerates all single-qubit stabilizer states (Pauli eigenstates) and projections (also referred to as effects) with their corresponding ZX-diagramatic representation. 
Many unitary gates also admit a simple description as ZX diagrams.
The Hadamard gate is already included as an elementary ZX tensor.
Arbitrary phase gates in the computational basis $Z_\alpha := e^{i\alpha/2} \cdot e^{-i \alpha Z/2}$ can also be represented through a single tensor node, as shown in Fig.~\ref{fig:zxintro1}c.
Similarly, a red spider with angle $\alpha$ and one input and output port corresponds to the operator $X_\alpha$.
Arbitrary single-qubit unitary gates can be constructed as products of these elementary gates.
The two-qubit entangling gates \texttt{CX} (also controlled-NOT or CNOT) in Fig.~\ref{fig:zxintro1}c and \texttt{CZ} in Fig.~\ref{fig:zxintro2}b provide illustrative examples of combining elementary ZX elements into known gates.
One may verify that the specified tensor contractions lead to operators which are proportional to the known two-qubit gates.
Each of the indices in a tensor can be reoriented from left to right (bra to ket) by contracting with the unnormalized reference state $\ket{\Omega} := \ket{00}+\ket{11}$ (respectively $\bra{\Omega} := \bra{00} + \bra{11}$); this is equivalent to \textit{partial matrix transposition}.
The reader may complain that vertical contractions as in Figs.~\ref{fig:zxintro1}c and \ref{fig:zxintro2}b are not well-defined as they do not specify which version of the tensors is used (i.e. a (2,1) spider or a (1,2) spider).
It turns out that in the case of ZX diagrams, any internal ambiguity in these options is irrelevant, which is a feature discussed in Appendix \ref{sec:PTinvariance}.

\textbf{ZX instruments.} It is not possible to describe fault tolerance without some amount of classical information being extracted from the quantum system.
At a high level, fault tolerance is the act of extracting the entropy introduced by noise out of the system by means of measurement (before it compromises logical degrees of freedom).
States and unitaries alone cannot describe this, as they lack a classical output.
We introduce an extension to the ZX-diagrammatic language to include \textbf{quantum instruments} (i.e., measurements and classical outcome generation) which provides enough expressive power to accurately represent stabilizer fault tolerance. 
We add classical outputs to spiders denoted by thick black lines, as shown in Fig.~\ref{fig:zxintro1}d. 
Each classical output is a bit $b$ (0 or 1) which determines whether the phase of the spider is $\alpha$ or $\alpha + \pi$. As shown in Fig.~\ref{fig:zxintro1}e, this can be used to express single-qubit measurements in terms of ZX instruments.
This amounts to projecting onto either one of two orthogonal stabilizer states in an outcome-dependent way.
In other words, for any fixed outcome configuration, we recover a plain vanilla ZX network.

Note that the instrument notation we use here differs from bastard spiders considered in Refs. \cite{Coecke2017}.
We elaborate on this difference in Appendix \ref{sec:NotationDiscrepancy}.

\textbf{Some ZX-diagrammatic equations.}
We provide a list of ZX-diagrammatic identities which we frequently find ourselves using when reasoning about circuits and fault-tolerance protocols in Fig.~\ref{fig:zxintro2}a.
We mostly use equations in a directed manner, which allows translating diagrams into a \textit{canonical form}.
This reflects our current goal of verifying the equivalence between different flavors of stabilizer fault tolerance.
Rigorous and general approaches to equivalence checking of ZX circuits exist \cite{Peham_2022} but are beyond our scope. The identities shown in Fig.~\ref{fig:zxintro2}a are:
\begin{itemize}

\item  {\bf Split and merge:} Two {\it simply connected} (via a single direct contraction)  like-colored spiders with phases $\alpha$ and $\beta$ can be merged into a single like-colored spider with the same set of outer ports and a phase $\alpha + \beta$.

\item {\bf Duality:} A red $\alpha$-spider is equivalent to a ZX network with a single green $\alpha$-spider with the same port signature and a Hadamard operator attached to each  port.

\item {\bf Identity:} A spider (green or red) with two ports and $\alpha=0$ (no phase annotation) is equivalent to a direct connector. 
This is the identity $\id$ as an operator, or $\ket{\Omega}$ as a state or $\bra{\Omega}$ as a projector. Consecutive pairs of Hadamard operators also equal the identity.
\end{itemize} 

Importantly, apart from the identity rule, these rules also work with the proposed extension to ZX {\bf instrument} networks.
The {\it duality} rule works by ignoring the classical output when attaching Hadamards.
The {\it merge} rule can also be generalized to the situation where one of the two like-colored spiders is an instrument. In this case, the resulting merged spider is itself a spider instrument with a single classical outcome. 
If two like-colored spider instruments are simply contracted, some care needs to be taken in order to apply the merge rule. A merge is possible, as long as only the joint parity of the outcomes is relevant for the composite instrument.
In this case, the merge will result in a single logical outcome bit which is the joint parity $b_1 \oplus b_2$ (the {\texttt{XOR}} of original outcome bits $b_1$ and $b_2$). 
As we are typically only concerned with the joint parity of the outcomes, the merge reduction summarizes the outcome information in exactly the right way.

{\bf Canonical form.} All red spiders can be removed and replaced by green spiders with the duality rule. 
The remaining rules can then be applied repeatedly to reduce the number of overall tensor nodes.
In this way, any ZX network can be recast in a unique canonical form.\footnote{By this, we do not mean that any two networks representing an equivalent tensor will reduce to the same representation, but rather, that given a starting ZX network, all reduction sequences which terminate will lead to the same final ZX network. 
This is the property known as {\it confluence} in abstract rewriting systems.}
This canonical form will have the following properties: There are no red spiders, no consecutive $H$ operators, no direct connections between green spiders (although 	they can be indirectly connected via an $H$), and all green spider have a degree different to 2 (unless $\alpha \neq 0$).

\textbf{Examples.} The example in Fig.~\ref{fig:zxintro2}b shows how a controlled-$Z$ gate can be expressed using two green spiders by applying the duality rule. Figure \ref{fig:zxintro2}c shows the example of a non-destructive two-qubit $Z \otimes Z$ measurement. 
In the circuit model, this can be expressed via an ancilla qubit initialized in the $|0\rangle$ state, two CNOT gates and a single-qubit $Z$ measurement. By expressing this as a ZX diagram and applying the transformation rules, we can reduce it to a diagram with three spiders. 
Finally, Fig.~\ref{fig:zxintro2}d shows a destructive Bell-basis measurement, i.e., a two-qubit measurement of the Pauli operators $Z \otimes Z$ and $X \otimes X$. There are two bits of classical information $b_{XX}$ and $b_{ZZ}$ which are extracted in a Bell measurement, such that, once the classical outputs are fixed, the ZX instrument in Fig.~\ref{fig:zxintro2}d represents one of four different ZX diagrams. 
Note that, while later examples will make use of the canonical form, those in Fig.~\ref{fig:zxintro2} are not in canonical form.

\begin{figure*}[t!]
	\centering
	\includegraphics[width=0.97\linewidth]{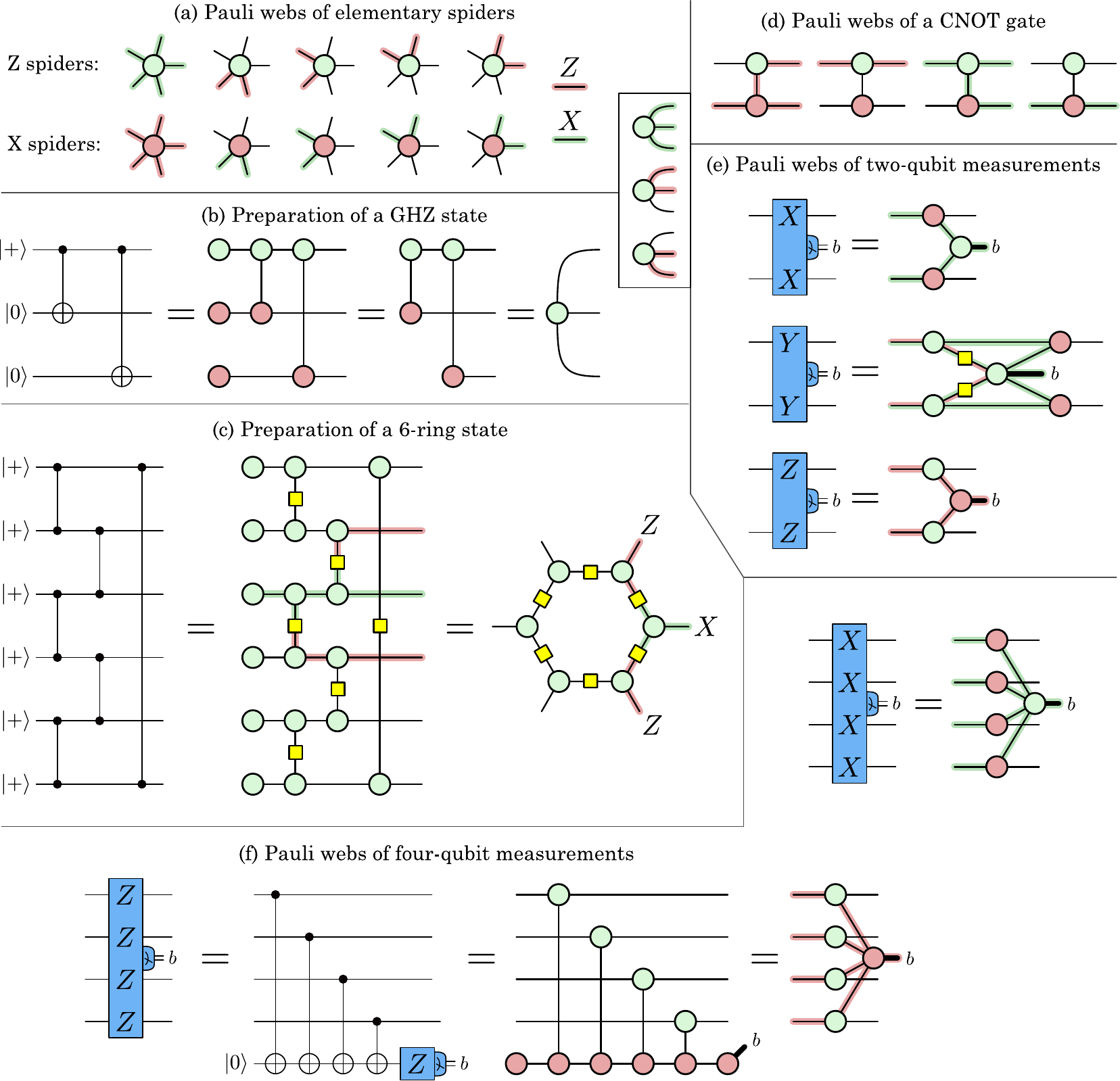}
	\caption{Pauli webs are a graphical overlay notation that allows us to reason about stabilizer states, Clifford gates and Pauli measurements. Pauli webs can describe the stabilizers of states, as shown for the example of a GHZ state (a/b) and a 6-ring graph state (c). They can also correspond to stabilizers of Clifford unitaries, as shown for the CNOT gate (d) and can specify Pauli projections and measurements, as shown for two-qubit (e) and four-qubit (f) Pauli measurements.}
	\label{fig:pauliwebs}
\end{figure*}

\section{Pauli webs}
Conveniently, spider nodes with $\alpha = 0$ (\textit{phaseless}) suffice to express most of stabilizer fault-tolerance.
This is because all resulting ZX instrument networks represent stabilizer instruments.
The entirety of a network can be efficiently described via a stabilizer representation (in addition to the instrument / tensor-network representation \cite{Bombin2021}).

We develop {\it Pauli webs} as a graphical overlay notation to identify the checks and stabilizers of ZX network diagrams.
Note that a similar notation has recently been used in Ref.~\cite{Gidney2023}.
We associate phaseless $n$-port spider nodes with $n$ stabilizer generators, $n-1$ of which are two-body stabilizer generators ($Z_iZ_{i+1}$ for green spiders and $X_iX_{i+1}$ for red spiders), and one additional $n$-body stabilizer generator ($X^{\otimes n}$ for green spiders and $Z^{\otimes n}$ for red spiders). 
Spiders with $\alpha = \pi$ or $\alpha=\pm \pi/2$ are also captured by the  stabilizer formalism. 
Specifically, green spiders with $\alpha = \pi$ have a $-X^{\otimes n}$ stabilizer instead of a $+X^{\otimes n}$, and red spiders $-Z^{\otimes n}$ instead of $+Z^{\otimes n}$. 
In the context of Pauli webs, $\pi$ spiders will support the same web structure as $0$-phase spiders apart from specific stabilizer presenting opposite sign.
In the context of stabilizer fault-tolerance, many of these signs will be determined at run-time based on actual measurement outcomes.
We will use a red and green highlight annotation, to emphasize the contraction/simplification of $Z$-type (red) and $X$-type (green) stabilizers. 
Figure \ref{fig:pauliwebs}a shows how this notation is used to highlight the stabilizer generators of single 5-port spiders.

Similar to how a single $n$-port spider is associated with $n$ stabilizer generators, a ZX tensor network containing multiple spiders and a total of $m$ unconnected ports (outer ports) is associated with $m$ stabilizer generators. 
These stabilizer generators can be found via overlay annotations that we refer to as Pauli webs.
In order to be consistent with the ZX network they are highlighting, valid Pauli webs need to have the following properties:
\begin{itemize}
	\item An even number of ports are highlighted red for any green spider.
	\item An even number of ports are highlighted green for any red spider.
	\item Either all ports or no ports are highlighted green (red) for any green (red) spider. 
	\item For spider instruments, if the classical outcome port is highlighted, then it must be highlight in the color of the spider.
	\item A Hadamard port is highlighted green if and only if its other port is highlighted red.	
\end{itemize}
These properties are designed to track the stabilizer nature of the ZX tensors.
Edges that are highlighted in both red and green are allowed by these rules.  The combination of two Pauli webs is also a valid Pauli web, where two overlapping highlights of the same color cancel.
If a valid Pauli web of a ZX network has highlighted outer ports, then the highlight for the outer ports represents a stabilizer of the network.
In fact, in any phase-free diagram with $n$ outer ports, there are $n$ independent Pauli webs with support on outer ports corresponding to $n$ stabilizer generators. There may be additional independent Pauli webs without support on outer ports corresponding to checks, as described later.

If a network contains $\pi$-phases and instrument outcomes, these may be required in order to identify the sign of the resulting network stabilizer.
The sign of the stabilizer is set by the parity of the number of $\pi$-phase spiders for which the $n$-body stabilizer is used.
If ZX instruments are involved, the sign is further corrected by the parity of the  highlighted classical outcomes.
Pauli webs allow us to use ZX diagrams to graphically reason about stabilizer states, Clifford gates, Pauli measurements and error-detecting checks, as we illustrate in the following examples.

\textbf{Stabilizer states.} 
If a ZX network is representing a state, the outer signature of Pauli webs describes the stabilizers of the state. 
The example in Fig.~\ref{fig:pauliwebs}b shows a circuit for the preparation of a 3-qubit GHZ state $(|000\rangle + |111\rangle)/\sqrt{2}$, which can be reduced to a single green 3-port spider. 
Since it has three ports, there are three independent Pauli webs, i.e., the single-spider Pauli webs specified in Fig.~\ref{fig:pauliwebs}a. 
These correspond to three independent stabilizer generators $Z \otimes Z \otimes \id$, $\id \otimes Z \otimes Z$ and $X \otimes X \otimes X$ of a GHZ state. 
The second example in Fig.~\ref{fig:pauliwebs}c shows the preparation circuit of a 6-qubit ring-shaped graph state, i.e., a 6-ring. 
The canonical form of the ZX diagram consists of 6 spiders, each contributing one outer port, so 6 independent Pauli webs can be found. The figure shows one of these Pauli webs tracking a ZX network transformation.
This highlights the natural one-to-one mapping between Pauli webs in ZX networks that are equivalent up to the transformations described in Fig.~\ref{fig:zxintro2}a.
For the 6-ring, the Pauli webs correspond to stabilizer generators of the form $Z_{j-1} \otimes X_j \otimes Z_{j+1}$ where the labeling of qubits follows cyclic ordering.

A general prescription for representing a graph state $\ket{G}$ corresponding to a graph $G$ uses the structure of the graph $G$ as a network.
Vertices in $G$ become green spider nodes with one more port than the \textit{degree} of the corresponding vertex. 
Edges of $G$ become Hadamard mediated tensor contractions among the corresponding nodes.
The prescription provided can be straightforwardly derived from the circuit construction of graph states by applying one {\texttt{CZ}} gate per edge of $G$ on the state $\ket{+}^{\otimes |G|}$.

\textbf{Clifford gates.} 
If a ZX network is representing a unitary Clifford gate $U$, the outer signature of Pauli webs specifies the gate as follows.
If a Pauli web for $U$ is supported with a Pauli operator $P_{in}$ on the input ports and $P_{out}$ on the output ports, this implies a map $U P_{in} U^\dagger = P_{out}$, i.e. $P_{in} \rightarrow P_{out}$. 
The example in Fig.~\ref{fig:pauliwebs}d shows the four generating Pauli webs of the ZX diagram of the CNOT gate (since it has four outer ports). If we label the ports as control ($c$) and target ($t$), as well as input ($in$) and output ($out$), then the four shown Pauli webs correspond to the stabilizers $Z_{t,in} Z_{c,out} Z_{t,out}$, $Z_{c,in} Z_{c, out}$, $X_{c, in} X_{c, out} X_{t, out}$ and $X_{t,in} X_{t,out}$. The implied Clifford operation then maps $Z_t \rightarrow Z_c Z_t$, $Z_c \rightarrow Z_c$, $X_c \rightarrow X_c X_t$ and $X_t \rightarrow X_t$, which is precisely the action of a CNOT gate.

\textbf{Pauli measurements.} 
If a ZX network is representing a projection, the outer signature of Pauli webs describes the Pauli basis of the projection.
If these Pauli webs also include classical outputs, then they can be thought of as multi-qubit Pauli measurements. 
The example in Fig.~\ref{fig:pauliwebs}e shows the two-qubit measurements $XX$, $YY$ and $ZZ$, where the Pauli webs associated with the measurements are highlighted.
Note that the Pauli web for the $YY$ measurement contains edges that are highlighted in both red and green.
In Fig.~\ref{fig:pauliwebs}f, constructions for the four-qubit $XXXX$ and $ZZZZ$ measurements are shown, with the corresponding Pauli webs highlighted.

\textbf{General Clifford operators.}
Stabilizer states, Clifford unitaries and stabilizer projections can be generalized to Clifford operators (see Appendix A of Ref. \cite{Bombin2021}).
These are operators $C$ which are specified by Pauli stabilizer equations of the form $P_{out} C P_{in} = C$ (up to a scalar).
Contrary to Clifford unitaries, general Clifford operators do not need to be unitary and may also include projections and isometries (such as an encoding circuit), as well as the examples mentioned earlier.

\begin{figure}
	\centering
	\includegraphics[width=\linewidth]{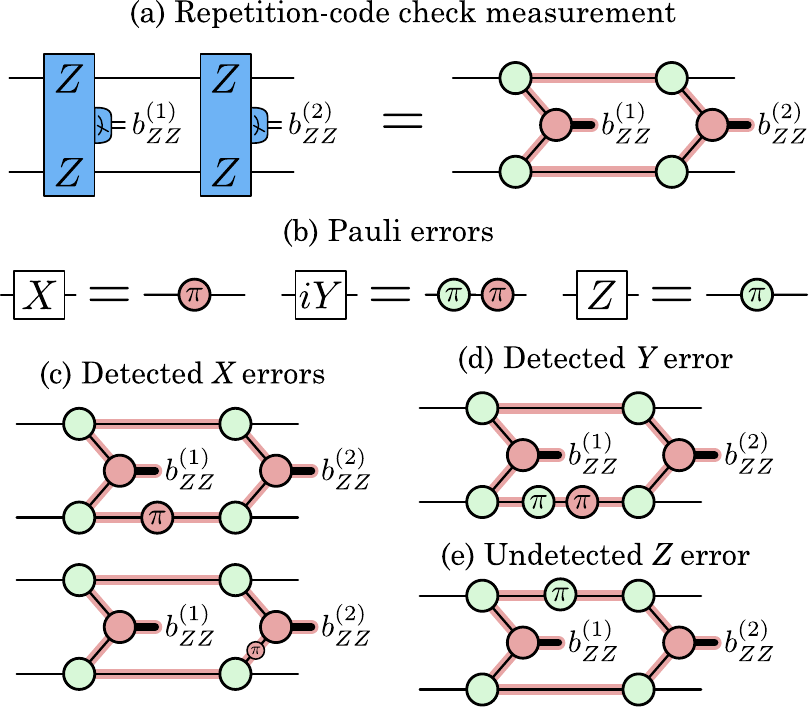}
	\caption{(a) Pauli webs that are completely internal (i.e., not supported on any output legs) describe checks, as shown for a repetition-code check. (b) Pauli errors can be described by inserting red and green spiders with a phase of $\pi$. (c) The repetition-code check can detect any odd number of $X$ error inserted along the Pauli web. (d) The same is true for $Y$ errors. (e) $Z$ errors cannot be detected, as the green spider has no effect on the red Pauli web.}
\label{fig:repcode}
\end{figure}

\textbf{Checks.} Finally, a ZX network may also contain Pauli webs that are not supported on any outer ports, i.e., have no outer signature and are completely internal. A non-trivial Pauli web involving instrument outcomes corresponds to a check.
They impose parity constraints on the outcome bits in the absence of errors.
In fault tolerance, checks are used to diagnose the presence and nature of errors in a physical implementation.
A simple example is the measurement of an error-detecting check in a two-qubit repetition code, as shown in Fig.~\ref{fig:repcode}a. Here, two ZZ measurements (as in Fig.~\ref{fig:zxintro2}c) in succession give rise to a check, as they must yield the same outcome in the absence of errors.
This check can be identified by a Pauli web in the corresponding ZX instrument network, as highlighted in Fig.~\ref{fig:repcode}a.
In this example, Pauli webs do not teach us anything we did not already know.
However, we have found them to be quite a useful book-keeping tool to understand more complex stabilizer computations.

The fact that the constituent tensors themselves are elementary stabilizer tensors brings the added benefit of {\it Pauli frame symmetry}.
Namely, the insertion of a $\pi$ spider node, either as an intentional operation or to represent an error, will not change the stabilizer \textit{structure} of the network. 
It will only flip the signs for a well defined set of check generators and network stabilizers.
Similarly, the post-selection upon any {\it valid} (compatible with checks) outcome configuration for ZX instruments, will lead to the same stabilizer structure (up to sign flips on a subset of generators).
This is precisely the effect that Pauli errors (or Pauli faults) have on a network. 
While we generally depict the fault-free representation of a quantum instrument network, it is possible to represent certain Pauli faults as the insertion of $X$, $Y$ or $Z$ operators at existing edges of the network (see Fig. \ref{fig:repcode}b).

This is a great modeling advantage of the network representation, which permits representing many error mechanisms which are physically motivated. The role of checks in fault tolerance is to make us aware of undesired faults in our physical implementation of a protocol.
Some of these faults can be represented by Pauli operator insertions within the edges of a network.
The examples in Fig.~\ref{fig:repcode}c-e represent how different faults/errors could give rise (or not) to a non-trivial check outcome (\textit{syndrome}). 
Figure \ref{fig:repcode}c shows that the insertion of an $X$ error in the circuit will be detected by the check, as the $\pi$ phase flips the product of measurement outcomes. 
In fact, the insertion of a red $\pi$ spider at any location along the red Pauli web will be detected, including measurement errors as shown in the bottom panel of Fig.~\ref{fig:repcode}c.
The same is true for $Y$ errors, as shown in Fig.~\ref{fig:repcode}d. 
However, $Z$ errors will not be detected, as green $\pi$ spiders have no effect on red Pauli webs, as shown in Fig.~\ref{fig:repcode}e. 
This is not surprising, since the repetition code is a classical error-correcting code, so it is only capable of detecting bit-flip errors, but not phase-flip errors.
Note that the per-edge error model implicit in the ZX network generally does not have a one-to-one correspondence to a circuit-level error model.

The example makes clear how only an odd number of detectable Pauli fault insertions will give rise to a non-trivial check syndrome.
This allows us to understand precisely what kind of errors each check is capturing and design them accordingly to maximize their usefulness. 
As we will see in the following section, a quantum error-correcting protocol such as the surface code is capable of detecting any low-weight combination of Pauli faults inserted in the ZX network, not just Pauli faults in one basis.

\begin{figure}[t!]
	\centering
	\includegraphics[width=\linewidth]{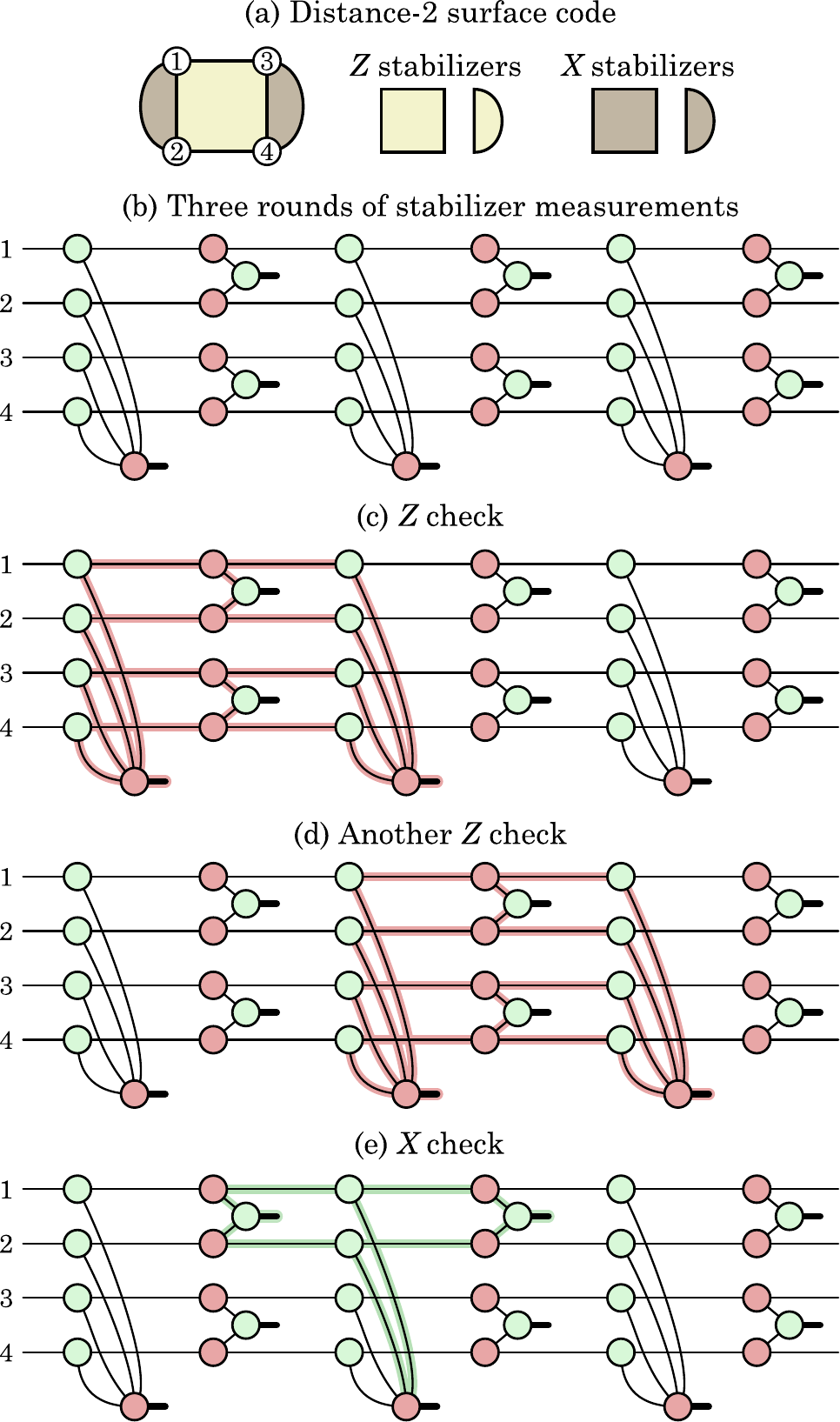}
	\caption{(a) A distance-2 surface code patch consisting of four physical qubits and three stabilizers. (b) ZX diagram of a circuit corresponding to three rounds of measurements of the three stabilizers. (c) Pauli web of a $Z$ check. (d) Pauli web of the subsequent $Z$ check. (e) Pauli web of an $X$ check.
	}
	\label{fig:surfacecode}
\end{figure}

\section{Expressing fault-tolerance as ZX instrument networks}\label{sec:ExpressingModels}

We have set up the ZX notation and reduction rules to express the equivalence of the different models of quantum computation.
In this section, we present (and simplify) each of the paradigms of stabilizer fault tolerance using the extended diagrammatic ZX notation. We focus on the example of surface codes.

\begin{figure*}[ht!]
	\centering
	\includegraphics[width=\linewidth]{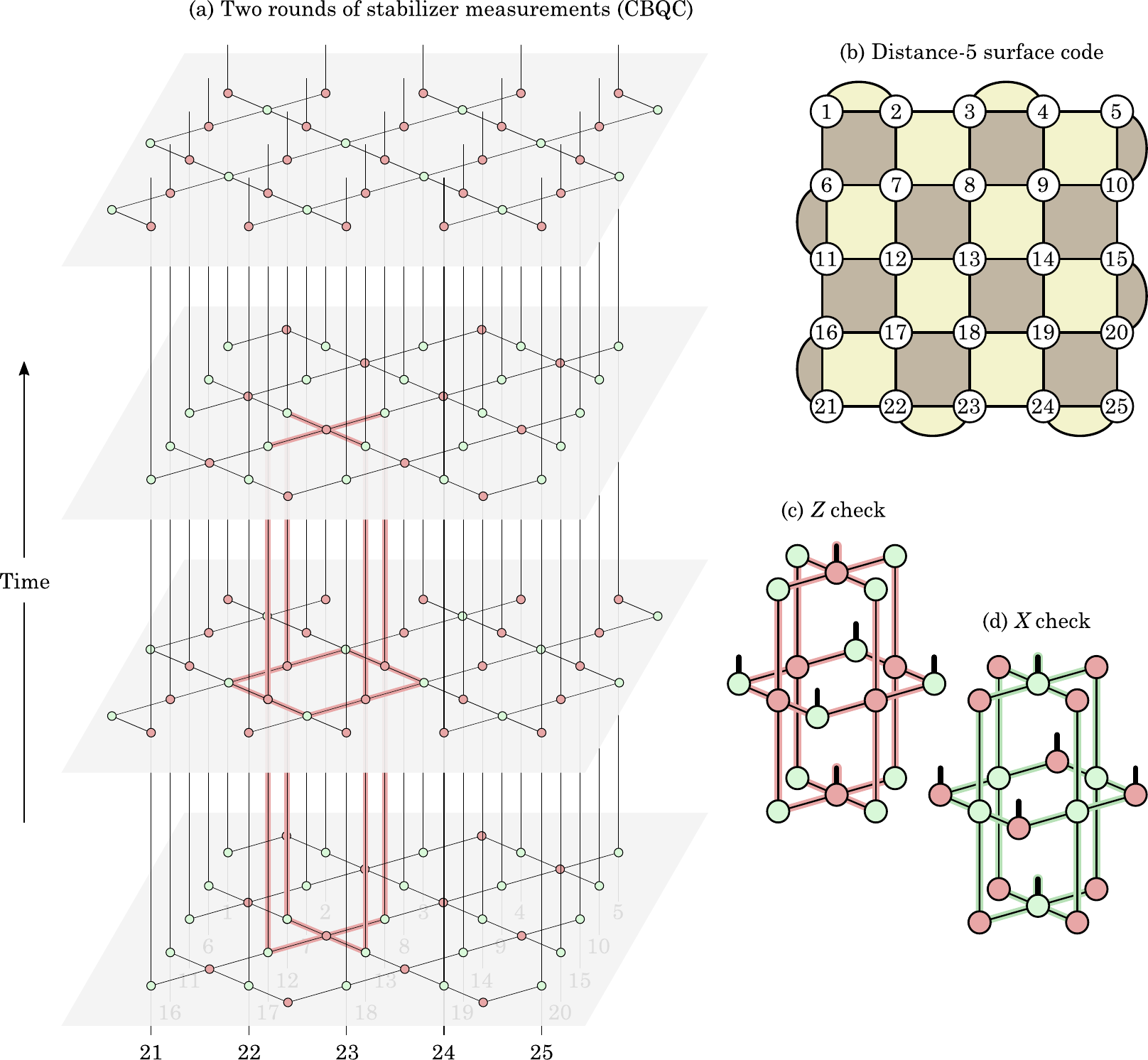}
	\caption{Circuit-based quantum computing. (a) Two rounds of the measurement of the 24 stabilizers shown in (b). Pauli webs of (c) $Z$ checks and (d) $X$ checks contain two measurement outcomes.}
	\label{fig:cbqc}
\end{figure*}

\begin{figure*}[ht!]
	\centering
	\includegraphics[width=\linewidth]{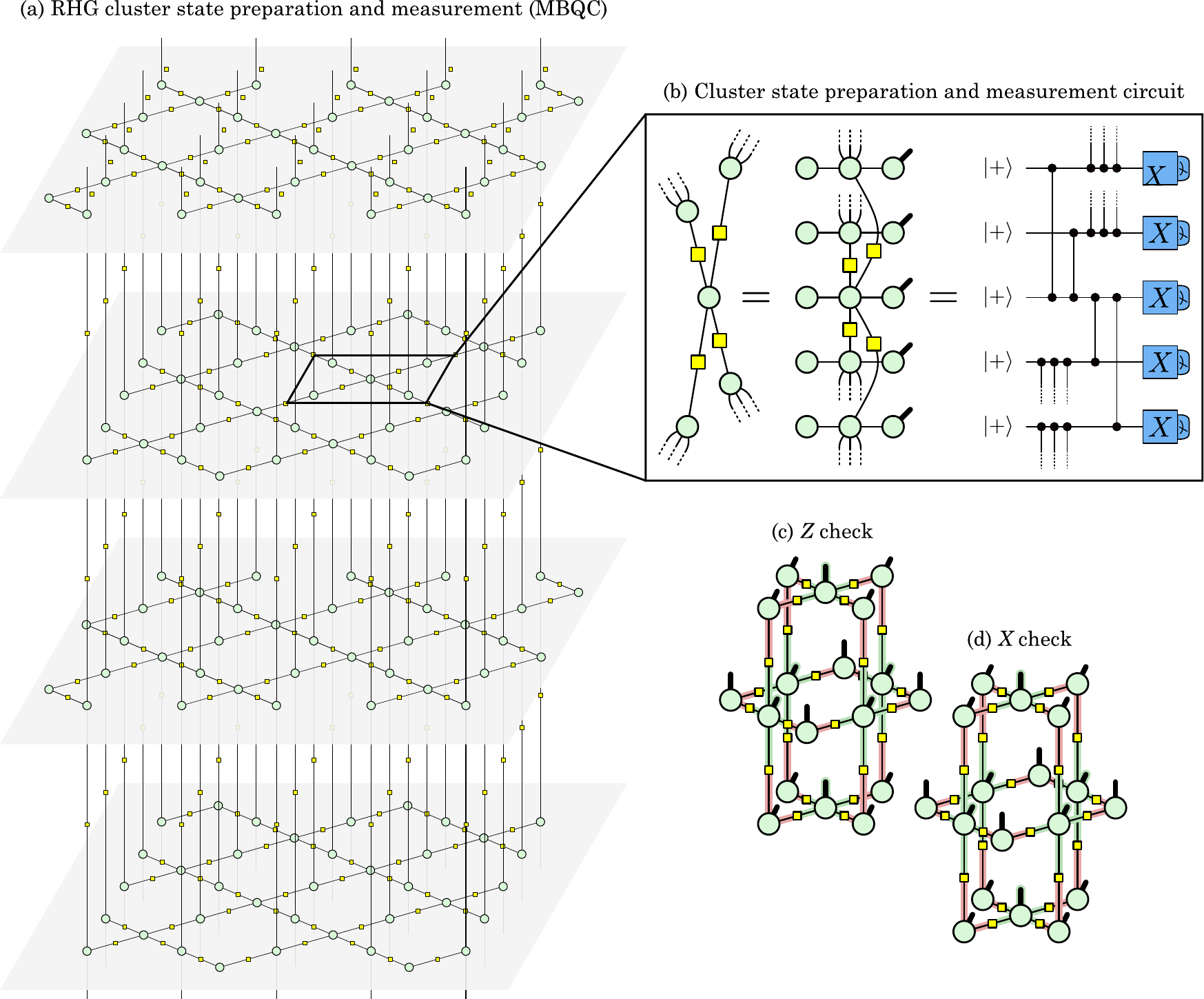}
	\caption{
		A common realization of fault-tolerant measurement-based quantum computing using a 3D cluster state. 
		(a) This lattice is obtained by applying the duality rule to the lattice in Fig.~\ref{fig:cbqc}a. 
		(b) It can be interpreted as the preparation of an RHG cluster state followed by  single-qubit $X$ measurements. 
		In other words, each green spider is replaced by a qubit initialized in the $|+\rangle$ state, followed by \texttt{CZ} gates between connected qubits and a single-qubit $X$ measurement. 
		(c) The $Z$ checks now contain six measurement outcomes. 
		(d) $X$ checks have the same structure as $Z$ checks, but are located in different layers.}
	\label{fig:mbqc}
\end{figure*}

\textbf{Circuit-based quantum computing (CBQC).} Having previously discussed the example of a two-qubit repetition code, we repeat the same analysis for a four-qubit surface code in Fig.~\ref{fig:surfacecode}. This code has three stabilizers $ZZZZ$, $XX\id \id$ and $\id \id XX$, as shown in Fig.~\ref{fig:surfacecode}a. The ZX diagram corresponding to a circuit containing three rounds of stabilizer measurements is shown in Fig.~\ref{fig:surfacecode}b, which consists of the two-body $XX$ measurements shown in Fig.~\ref{fig:pauliwebs}e and four-body $ZZZZ$ measurements in Fig.~\ref{fig:pauliwebs}f. We can identify $Z$ checks as internal Pauli webs consisting of pairs of consecutive measurement outcomes, see Figs.~\ref{fig:surfacecode}c and d. However, in contrast to repetition codes, we now also find green internal Pauli webs as $X$ checks, as shown in Fig.~\ref{fig:surfacecode}e. 
The check structure implies that any single-edge Pauli error inserted into the network can be detected by the checks, as each edge contributes a red segment to at least one internal Pauli web and a green segment to at least one other internal Pauli web.

\begin{figure*}[t!]
	\centering
	\includegraphics[width=0.95\linewidth]{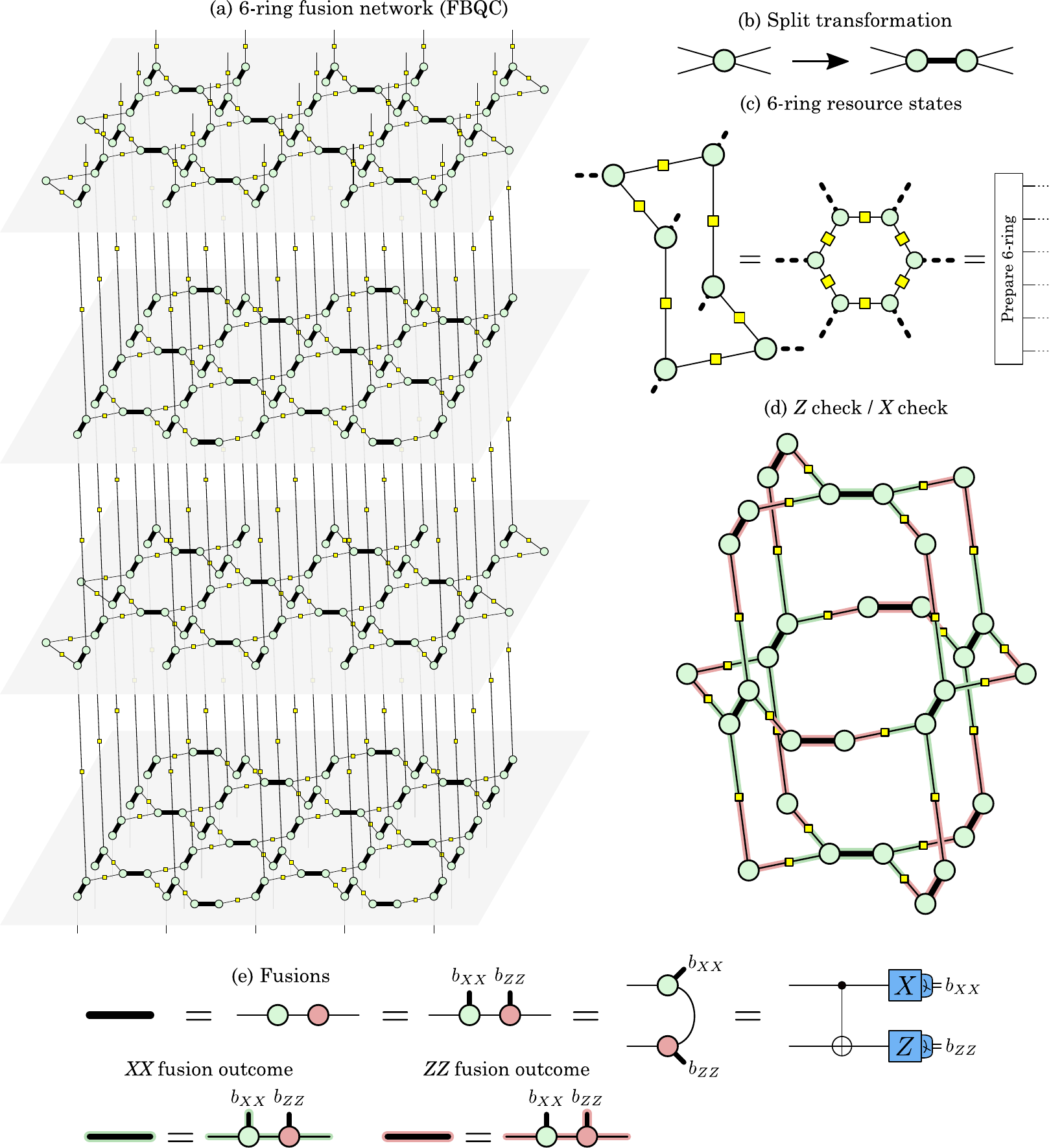}
	\caption{Fusion-based quantum computing. (a) This lattice is obtained by applying the split transformation to the lattice in Fig.~\ref{fig:mbqc}a. The thick lines partition the lattice into many copies of six-ring resource states. 
		The ZX diagram can then be interpreted as the preparation of many copies of 6-ring resource states (c), followed by their fusion 
		(e), i.e., the destructive two-qubit measurement of pairs of qubits from different resource states. 
		(d) Each check contains 12 fusion outcome bits.}
	\label{fig:fbqc}
\end{figure*}

Next, we consider a distance-5 surface-code patch consisting of 25 physical data qubits in Fig.~\ref{fig:cbqc}. 
Since the qubits are arranged on a 2D square grid, and stabilizer measurements are local operations, it is convenient to draw the circuit as a three-dimensional object. 
If we straightforwardly translate the circuit  corresponding to two rounds of stabilizer measurements into a ZX diagram, we obtain the 3D diagram shown in Fig.~\ref{fig:cbqc}a, where the time direction goes from bottom to top, rather than left to right as in previous circuit diagrams. 
The first layer in this diagram contains 12 two-body and four-body $Z$ measurements, and the second layer 12 two-body and four-body $X$ measurements. These measurements are repeated in the third and fourth layers. 
We can identify checks as internal Pauli webs that form three-dimensional cells, one of which is highlighted in Fig.~\ref{fig:cbqc}a. These checks have the structure shown in Fig.~\ref{fig:cbqc}c, consisting of red Pauli webs containing pairs of consecutive measurement outcomes. Similarly, $X$ checks consist of green Pauli webs, as shown in Fig.~\ref{fig:cbqc}d. These are the checks that are used to detect and correct errors in a surface-code decoder.

\begin{figure*}[t!]
	\centering
	\includegraphics[width=\linewidth]{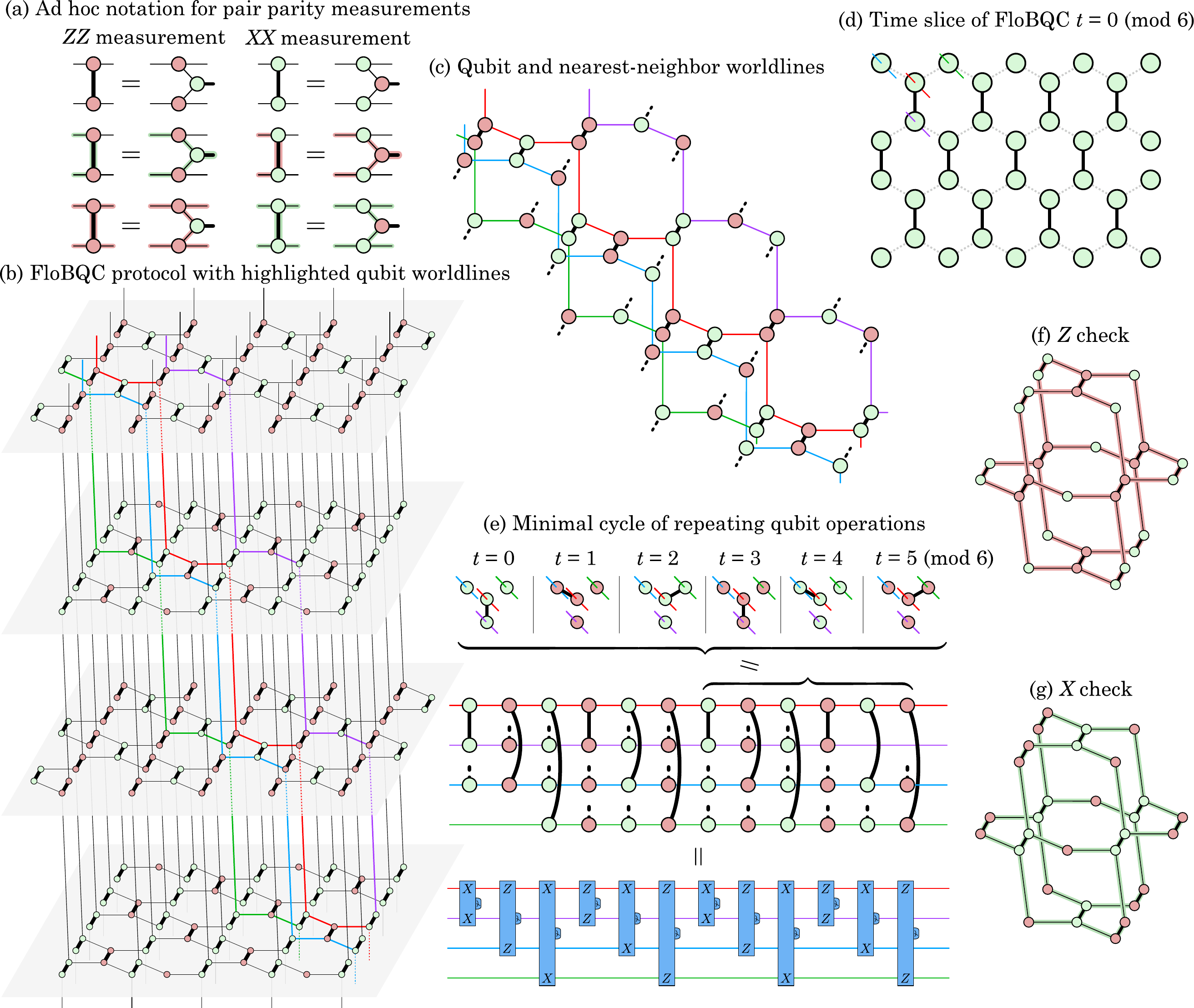}
	\caption{Floquet surface code. (a) Local to this figure we take an ad hoc interpretation of thick black lines connecting like colored spiders as a pair parity measurement.
	(b) The FloBQC protocol which is obtained by applying a split transformation to the lattice in Fig.~\ref{fig:cbqc}a. 
	The thick black connections partition the lattice into individual chains, four of which are highlighted in red and its neighbors green, blue and purple. 
	This is equivalent to the Floquet code presented in Refs.~\cite{Davydova2023} and~\cite{Kesselring2022}.
	(c) These chains can be interpreted as four qubits that undergo a cycle of six two-body measurements, alternating between $XX$ and $ZZ$ measurements and cycling through the three neighbors of each qubit. 
	(d) The physical qubits form a hexagonal lattice when taking a time slice.
	Pair parity measurements of the same type and along the same orientation are preformed simultaneously.
	(e) The pair parity measurements alternate between $ZZ$ and $XX$ measurements and cycle along the three neighbors of each qubit leading to an overall cycle of six operations.
	The $Z$ checks (f) and $X$ checks (g) consist of six $ZZ$ and $XX$ measurement outcomes, respectively.}
	\label{fig:flobqcV2}
\end{figure*}

\textbf{Measurement-based quantum computing (MBQC).} We can construct different models of quantum computation by reinterpreting this three-dimensional ZX diagram. One interpretation is obtained by putting the diagram into its canonical form by applying the duality rule, as shown in Fig.~\ref{fig:mbqc}a. The resulting diagram is a network of green spiders, where each spider in the bulk is connected to four other green spiders via a Hadamard gate. As shown in Fig.~\ref{fig:mbqc}b, we can interpret each such spider as a qubit that is prepared in the $|+\rangle$ state, participates in four \texttt{CZ} gates with its four neighbors, and is then measured in the $X$ basis. 
The state that is prepared and then measured is an RHG cluster state~\cite{raussendorf2006} (or RBH state~\cite{Raussendorf2005}), which is a graph state similar to the example shown in Fig.~\ref{fig:pauliwebs}c. 

The checks have a similar structure as in the CBQC protocol, but now consist of six measurement outcomes, as shown in Figs.~\ref{fig:mbqc}c and d. Note that $Z$ checks and $X$ checks have an identical structure, differing only in their location. The cells centered around every second layer are $Z$ checks, whereas the cells centered around every other layer are $X$ checks. $Z$ and $X$ checks are also referred to as primal and dual checks.

Interestingly, the resulting canonical form has no specific orientation which should naturally be associated to a time-like direction.
By applying the merge rule to the circuit preparing the cluster state, we have compacted the time-like direction associated with the cluster-state preparation circuit.
It is nevertheless possible to reintroduce a time-like ordering (direction) to the diagram, which in a circuit interpretation allows reducing the number of physical qubits which need to be present at any given time.
In other words, state preparation and consumption (measurement) may proceed concurrently on different parts of the resource state so long as preparation precedes measurement for any given part.

The ZX network thus described is locally bipartite (Hadamard nodes are not counted).
This allows absorbing the Hadamard operators into the nodes in one part of the bi-partition  and having them change color through the duality identity.
The fact that we have a choice on which half of the spider instruments to swap color is a reminder that the separation into primal and dual checks is an arbitrary choice.
Moreover, if the lattice is not bipartite but only \textit{locally bipartite} then the network is not bicolorable and it is not possible to remove all Hadamard operators through the duality transformation.
This implies that there is no locally consistent separation into primal and dual checks. 
A \textit{transparent boundary}~\cite{Bombin2021}, which is where these labels swap,  must be introduced along some cut interrupting all odd loops. 

Other cluster-state geometries are possible following this recipe. For example, using the fault-tolerant cluster states of Ref.~\cite{Nickerson2018}, which can be defined for any cell-complex (in any dimension), one can replace qubits residing on one type of cell with red spiders, and the other with green spiders, and connect them up appropriately. 

\begin{figure*}[t!]
	\centering
	\includegraphics[width=\linewidth]{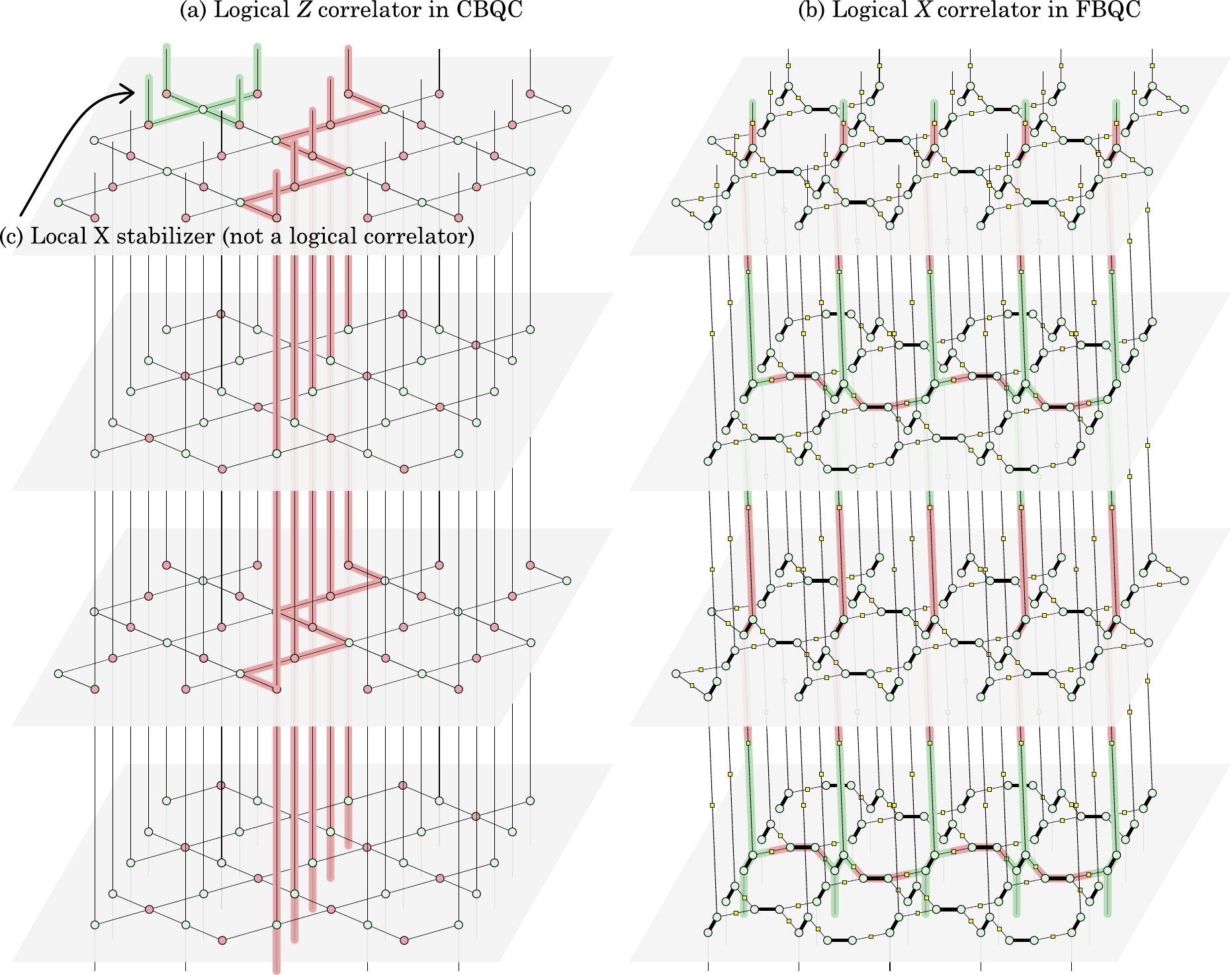}
	\caption{(a) A logical $Z$ correlator highlighted in the lattice of Fig.~\ref{fig:cbqc}a. The Pauli web connects the logical $Z$ operator on the inputs to the logical $Z$ operator on the outputs. (b) A logical $X$ correlator highlighted in the lattice of Fig.~\ref{fig:fbqc}a, connecting the logical $X$ operators on the inputs and outputs. (c) A Pauli web supported on outer ports corresponding to a local stabilizer instead of a logical correlator. This local stabilizer can be completed into a check by continuing the fault-tolerant protocol.}
	\label{fig:correlators}
\end{figure*}

\textbf{Fusion-based quantum computing (FBQC).}
An illuminating way to view the concrete FBQC protocol presented in Ref.~\cite{bartolucci2021} is as a factorization of the canonical ZX network diagram into resource-state-sized tensor factors (disregarding the instrument nature of nodes).
One possibility to do so is to apply a split to every spider node of the diagram in Fig.~\ref{fig:mbqc}a, i.e., applying the transformation shown in Fig.~\ref{fig:fbqc}b. This results in the diagram shown in Fig.~\ref{fig:fbqc}a, where the new edges introduced via the split transformation are highlighted as thick black lines. Note that if the lattice is cut along these thick black lines, it is partitioned into individual pieces each containing six spiders in the bulk. These pieces form hexagonal rings, as shown in Fig.~\ref{fig:fbqc}c. We can interpret these pieces as circuits preparing copies of 6-ring graph states, which we refer to as \textit{resource states}.

Furthermore, we can interpret the thick black lines connecting pairs of different resource states as two-qubit Bell measurements, which we refer to as \textit{fusions}, as shown in Fig.~\ref{fig:fbqc}e.
In an implementation with linear optics, it will, in practice, be a locally encoded fusion measurement, which is required in order to increase the entangling probability. 
These fusions produce two classical outcomes, which we will highlight as red and green Pauli webs around the thick black line. The internal Pauli webs corresponding to surface-code checks have a similar structure as in previous examples, but now consist of 12 fusion outcomes, six of which are $XX$ outcomes, and six $ZZ$ outcomes (see Fig.~\ref{fig:fbqc}d). Note that the ZX diagram does not specify the order in which resource states need to be generated and fused. They can be generated all at once for a fast but hardware-intensive computation, layer by layer, or, in the extreme case, completely sequentially via a technique referred to as \textit{interleaving}~\cite{Bombin2021a}.

As in the MBQC case, the ZX network is, strictly speaking,  not equivalent to the CBQC network in Fig.~\ref{fig:cbqc}a because there is a larger and different  set of classical outcomes attached to the network.
However, because the Bell measurement is composed of two-port spider instruments without phases, it  just contributes an outcome-dependent Pauli frame.
One can nevertheless identify the Pauli webs for check generators from FBQC with the check generators in the CBQC picture by applying the transformations onto the MBQC canonical form.

In general, there may also be more outcome bits which determine the Pauli frame correction associated to the resource state generation.
This Pauli frame correction can be specified by using at most as many bits as the number of stabilizers.
For graph states, it is conventional to take these bits in one to one correspondence with the sign correction of vertex stabilizers (which form a generating set).

\textbf{A Floquet model (FloBQC).}
Floquet-flavor fault-tolerance is adapted to a physical setting where the native instructions are non-destructive two-body measurements.
Our initial CBQC diagram in Fig.~\ref{fig:cbqc}a can also be used to obtain a Floquet flavor fault-tolerance protocol. We apply a split to each spider to obtain the lattice in Fig.~\ref{fig:flobqcV2}a. 
Note that this split is different than in Fig.~\ref{fig:fbqc}, since cuts along the thick black lines break the lattice into long disjoint chains rather than hexagons.
Four of these chains are highlighted in red, green, blue and purple. 
The thick black lines connect each chain to three other chains, in this case the red chain to its three nearest neighbors, i.e., the green, blue and purple chains. 
We can interpret these chains as qubits in a circuit, and connections between chains as two-body $XX$ or $ZZ$ measurements, as shown in Fig.~\ref{fig:flobqcV2}b.

In this case, we recover the FloBQC protocol from Refs.~\cite{Davydova2023} and \cite{Kesselring2022}. 
Each chain is associated with a physical qubit. 
Each physical qubit has three neighbors and they are arranged in a honeycomb lattice.
The fault tolerance protocol alternates between $XX$ and $ZZ$ measurements, cycling between the three neighbors, resulting in an overall cycle of six two-body measurements, as shown in Fig.~\ref{fig:flobqcV2}. Checks are obtained as products of six measurement outcomes, as shown in Figs.~\ref{fig:flobqcV2}c and d. Note that only Pauli webs of one color (but not the other) are attached to classical outcomes on the thick black connections.

It is interesting to note, that both CBQC and FloBQC have a notion of physical qubits with identities persisting through time and subjected to a sequence of elementary operations dictated by the fault-tolerance protocol.
However, the direction in which the strands representing such physical qubits propagate are different in the CBQC model and the FloBQC when mapped to the shared ZX-network (see Fig. \ref{fig:summary})).
This is consistent, as the logical correlator membranes can be deformed and reoriented and the true direction of encoded information flow is determined by the arrangement of boundary features.
We have only cursorily introduced these for the circuit model, and leave the industrious reader to reinterpret boundary conditions in the context of the different flavors of fault-tolerance.
Identifying the mapping through a common simplified  ZX diagram is particularly useful for this.

It is also worth pointing out that, as with all ZX networks, the interpretation as a circuit-based protocol is only one of many possibilities. 
For instance, the same network can also be interpreted as an FBQC protocol with linear resource states.

\textbf{Logical correlators.} So far, our main focus was on describing checks as internal Pauli webs. The 3D ZX diagrams also contain Pauli webs with support on the outer ports. Some of these Pauli webs correspond to stabilizers that are local to each logical port, an example of which is shown in Fig.~\ref{fig:correlators}c. Other such Pauli webs correspond to logical correlators (also called logical membranes or correlation surfaces) which encode the correlation between logical input observables and logical output observables. The logical operations in our examples are simple idle operations, so the membranes connect the logical $Z$ and $X$ operators at the inputs to the logical $Z$ and $X$ operators at the outputs. An example of a logical $Z$ correlator in the CBQC lattice is shown in Fig.~\ref{fig:correlators}a. The Pauli web is supported on the $Z^{\otimes 5}$ operators corresponding to representatives of the logical $Z$ operator on the input and output. 

Note that the Pauli webs of logical correlators in CBQC do not involve any classical outcomes. 
However, we still need to keep track of a Pauli frame in CBQC to account for Pauli errors on the edges of logical correlators. The locations of these errors are diagnosed by the decoder. 
In contrast, the logical $X$ correlator in the FBQC lattice shown in Fig.~\ref{fig:correlators}b is given by a Pauli web which involves many different fusion outcomes. This is also the case for logical correlators in MBQC and FloBQC. 
In these models of computation, a Pauli frame is required even in a noiseless setting, as logical correlators contain intrinsically random measurement outcomes in addition to edges that may host Pauli errors.

\textbf{Fault-tolerant logical operations.}
Using the equivalences provided by the ZX tensor networks, we have shown that the different models share a common structure. This provides a recipe for constructing fault-tolerant logical operations in different models of computation. In particular, one can use the prescriptions in Ref.~\cite{Bombin2021} to construct topological features (boundaries, domain walls, symmetry defects) realizing fault-tolerant quantum instruments~--~called \textit{logical blocks}~--~in the different models.

\section{Conclusion}\label{sec:conclusions}

\begin{figure}[t!]
	\centering
	\includegraphics[width=\linewidth]{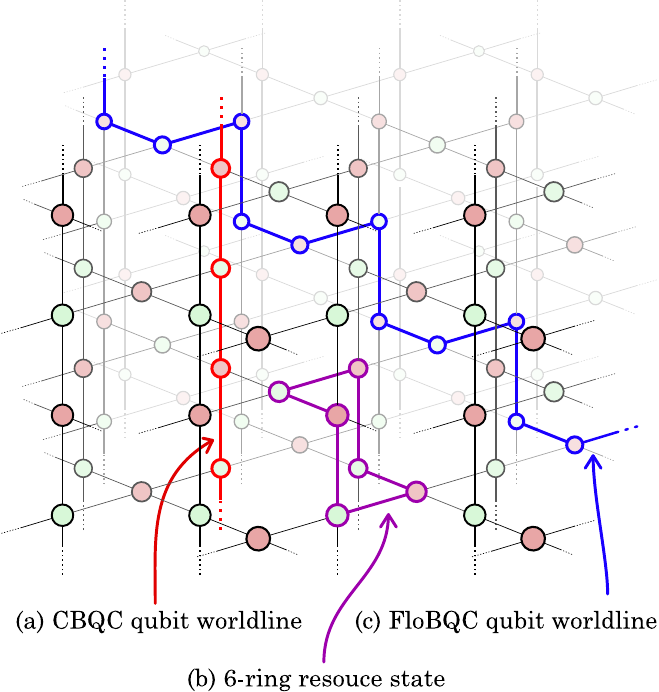}
	\caption{Different models of quantum computation shown in the same ZX network.
Note that the canonical form of this network can be obtained by applying the duality rule to red spiders.
It emphasizes the equivalence of all three Cartesian axes in the space-time picture.
(a) A worldline for a physical qubit in the traditional circuit model is highlighted in red.
(b) A resource state for the FBQC scheme presented in Ref.~\cite{bartolucci2021} is highlighted in purple.
(c) A worldline for a FloBQC physical qubit is highlighted in blue and propagates in a different direction compared to a CBQC qubit.
Note that, in this figure, each ZX spider is shared by two resource states in FBQC or by two qubits in FloBQC. In other words, a split transformation needs to be applied to obtain the networks in Figs.~\ref{fig:fbqc} and \ref{fig:flobqcV2}. }
	\label{fig:summary}
\end{figure}

Throughout this article we have presented  how the different paradigms of stabilizer fault tolerance share a common structure.
Of course, saying that they are the same would be overly exaggerated.
Each has its own set of outcomes produced in different ways and the error models adequate for describing each platform will differ.
Nevertheless, the structure of checks and logical correlators is shared and the examples provided in this article illustrate how to translate between paradigms.

The models are equivalent in the sense that they can be reduced to a common canonical form via local ZX transformations which preserve the check structure, as illustrated in Fig.~\ref{fig:summary}. In particular, the model of FloBQC that we present, while conceptually different, is ultimately based on the same underlying surface code. As such, one can define logical operations in the same manner as in other flavors of surface-code quantum computation.

While this paper focuses on the bulk of the most basic topological protocol, it should be interpreted as a dictionary.
We hope that fault-tolerance researchers will find it useful for the translation between paradigms beyond the previous examples. This may include different microscopic models (i.e., crystal structures~\cite{Nickerson2018,Newman2020}), features (boundary conditions, logical blocks~\cite{Bombin2021}, transversal gates, etc.), as well as different fault-tolerance protocols based on color codes~\cite{Bombin2006,Bombin2018}, low-density parity check codes~\cite{Breuckmann2021} or other Clifford encoders~\cite{Khesin2023}. We found the ZX calculus to be a versatile toolbox that can be used at all levels of fault tolerance, from the physical level of different models of quantum computation, to the structure of checks used in decoding~\cite{Bombin2023}, and the methodical construction of logical operations~\cite{Litinski2022}.

\subsection*{Author contributions}
HB derived the version of FloBQC described in this paper through equivalence with FBQC before these protocols were known as Floquet surface codes in the literature.
FP developed the Pauli web notation for visually tracking Pauli stabilizers and identifying checks,
as well as the notation for ZX instruments with classical outputs used in this paper.
HB, DL, NN, FP and SR contributed to the unification of fault tolerance methods between CBQC, MBQC and FBQC.
DL and FP wrote the manuscript. 
HB requested that this acknowledgment include that he has not reviewed the paper in its final form.

\subsection*{Acknowledgments}
We thank thank Daniel Barter, Ye-Hua Liu, Sam Morley-Short and Terry Rudolph for useful comments on an earlier draft. Many of the ideas presented in this manuscript were developed through collaborations with our colleagues at the PsiQuantum architecture team.

\appendix

\section{Importance of permutation and partial transposition invariance in elementary ZX tensors}\label{sec:PTinvariance}

\begin{figure}[h!]
	\centering
	\includegraphics[width=\linewidth]{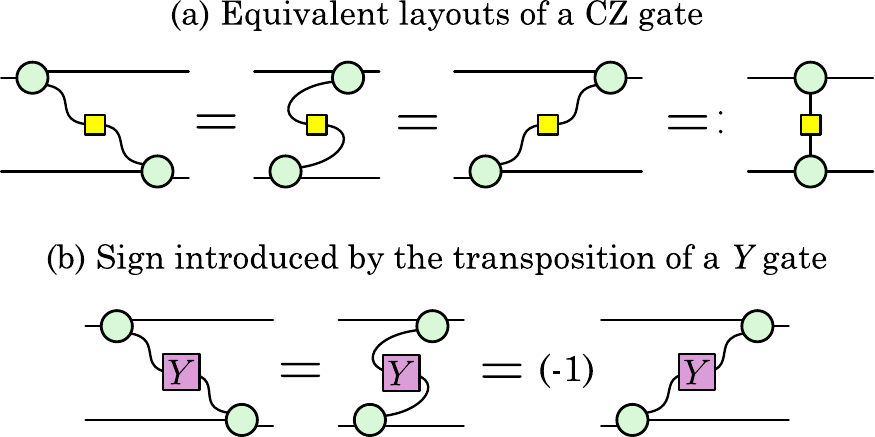}
	\caption{
		(a) Both spider tensor definitions and Hadamard are compatible with index permutation and partial/full transposition.
		This means that all topologically equivalent ways of presenting ZX networks are equivalent (up to identification of outer ports).
		This allows making sense of a vertical tensor contraction, as a slight reorientation of the contraction in either direction would give rise to the same result. 
		(b) This feature requires all elementary tensors to be invariant under permutations and transpositions and is not valid if we, e.g., include $Y$ or $e^{i \alpha Y}$ as an elementary tensor.
		Transposing contractions in a diagram containing $Y$ operators may result in minus signs.
	}
	\label{fig:4spider-to-projective-meassurement}
\end{figure}

A feature which makes ZX diagrams particularly light-weight and easy to use is that it is not necessary to distinguish ports in the  elementary tensors of the diagram.
This is because all elementary tensor definitions are  invariant under index permutations and compatible with partial transposition.
This allows the network to be interpreted in a way which only depends on the connectivity between nodes and where it does not matter if a port emerges in the forward direction (a ket) or in the backwards direction (a bra).
For instance, the transpose of the $Y$ operator is its negative $(Y^T = -Y)$, as shown in Fig.~\ref{fig:4spider-to-projective-meassurement}b. 
Introducing any such tensor would require specifying its orientation in every diagram it is involved.
This is the case in other tensor network diagrams; one needs to specify the signature of each tensor (how many kets and how many bra indices), and specify to which port it is being connected (if there is a tensor with multiple qubits). 
Avoiding this overhead makes ZX diagrams significantly lighter in notation while retaining rigorous meaning.

Note that the results of composite networks do not need to have permutation invariance or partial transposition invariance, but they will do so if the the diagram itself is invariant under permutations of outer ports. An example is the controlled $Z$ gate shown in Fig.~\ref{fig:4spider-to-projective-meassurement}a, which is invariant under transposition and commutes with the SWAP operator.

The extension of ZX diagrams to instruments presented in the main text requires additional instrument nodes.
We chose these instrument nodes such that the properties of permutation and partial transposition invariance continue to hold for the quantum ports in all elementary instruments.
This does not include classical outcome ports, as they are treated separately.

\section{Notation discrepancy with literature \label{sec:NotationDiscrepancy}}
As pointed out to us by Bob Coecke, the notation for ZX instruments used in this article  differs from the one established in existing literature.
We interpret thick edges,  introduced in Fig.~\ref{fig:zxintro1}d, as classical outcomes or control wires with further ad hoc uses in Figs.~\ref{fig:fbqc} and \ref{fig:flobqcV2}.
In contrast, books such as Ref. \cite{Coecke2017} use unpaired simple wires to represent classical states and outcomes, whereas thick (or double) edges  are used as a shorthand for doubled wires, which carry the bra and ket components of quantum states.

\begin{figure}[h!]
	\centering
	\includegraphics[width=\linewidth]{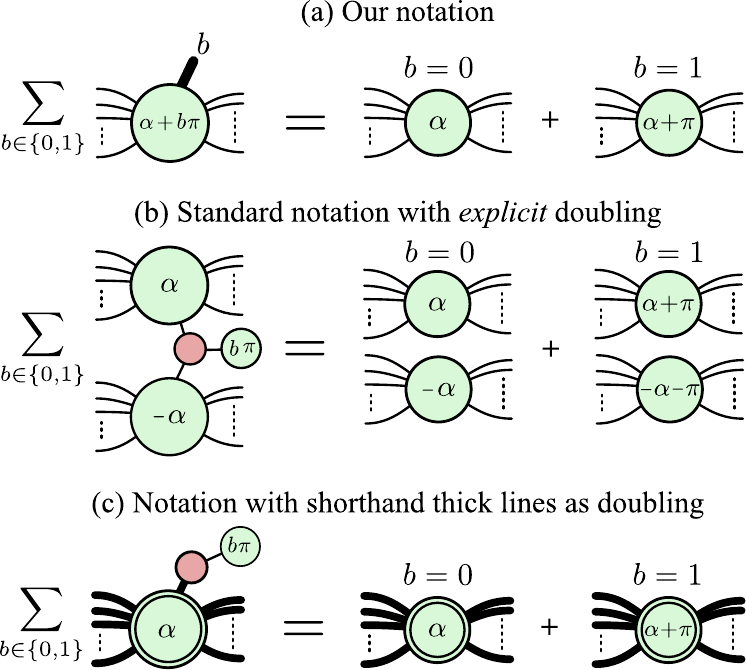}
	\caption{(a) A green spider instrument is presented in the notation of this article.
		(b) Without use of thick edges the doubling of bra and ket components is done explicitly.
		(c) The same instrument is now expanded in a notation where thick edges represent quantum wires (i.e. doubled wires) and double lined spiders represent pairs of spiders with opposite phase.
	}
	\label{fig:notationclarification}
\end{figure}
Our notation can be interpreted as a shorthand for the standard notation, as shown in Fig.~\ref{fig:notationclarification}. 
In such a case, the identities involving quantum instruments in Fig.~\ref{fig:zxintro2}  may be derived from elementary ZX identities and do not need to be postulated.

While succinct, the notation we introduce is sufficient to represent stabilizer fault tolerance.
Furthermore, having classical outcomes and inputs be explicitly represented as ports in the diagram provides a clear type signature for the classical control which should accompany the quantum computer, which we do not expect to represent through ZX diagrams.

\bibliographystyle{quantum}

\begin{thebibliography}{36}	
	\bibitem{kitaev2003}
	Alexei~Yu Kitaev.
	\newblock ``Fault-tolerant quantum computation by anyons''.
	\newblock \href{https://dx.doi.org/10.1016/S0003-4916(02)00018-0}{Annals of Physics {\bf 303}, 2--30}~(2003).
	
	\bibitem{dennis2002}
	Eric Dennis, Alexei Kitaev, Andrew Landahl, and John Preskill.
	\newblock ``Topological quantum memory''.
	\newblock \href{https://dx.doi.org/10.1063/1.1499754}{Journal of Mathematical Physics {\bf 43}, 4452--4505}~(2002).
	
	\bibitem{raussendorf2003}
	Robert Raussendorf, Daniel~E. Browne, and Hans~J. Briegel.
	\newblock ``Measurement-based quantum computation on cluster states''.
	\newblock \href{https://dx.doi.org/10.1103/PhysRevA.68.022312}{Physical Review A {\bf 68}, 022312}~(2003).
	
	\bibitem{briegel2009}
	H.~J. Briegel, D.~E. Browne, W.~Dür, R.~Raussendorf, and M.~Van~Den Nest.
	\newblock ``Measurement-based quantum computation''.
	\newblock \href{https://dx.doi.org/10.1038/nphys1157}{Nature Physics 2009 5:1 {\bf 5}, 19--26}~(2009).
	
	\bibitem{raussendorf2006}
	R.~Raussendorf, J.~Harrington, and K.~Goyal.
	\newblock ``A fault-tolerant one-way quantum computer''.
	\newblock \href{https://dx.doi.org/10.1016/j.aop.2006.01.012}{Annals of Physics	{\bf 321}, 2242--2270}~(2006).
	
	\bibitem{bartolucci2021}
	Sara Bartolucci, Patrick Birchall, Hector Bombin, Hugo Cable, Chris Dawson,
	Mercedes Gimeno-Segovia, Eric Johnston, Konrad Kieling, Naomi Nickerson,
	Mihir Pant, et~al.
	\newblock ``Fusion-based quantum computation''.
	\newblock \href{https://dx.doi.org/10.1038/s41467-023-36493-1}{Nature Communications {\bf 14}, 912}~(2023).
	
	\bibitem{hastings_2021}
	Matthew~B. Hastings and Jeongwan Haah.
	\newblock ``Dynamically generated logical qubits''.
	\newblock \href{https://dx.doi.org/10.22331/q-2021-10-19-564}{Quantum {\bf 5}, 564}~(2021).
	
	\bibitem{paetznick2022}
	Adam Paetznick, Christina Knapp, Nicolas Delfosse, Bela Bauer, Jeongwan Haah,
	Matthew~B. Hastings, and Marcus~P. da~Silva.
	\newblock ``Performance of planar {F}loquet codes with {M}ajorana-based
	qubits''.
	\newblock \href{https://dx.doi.org/10.1103/PRXQuantum.4.010310}{PRX Quantum {\bf 4}, 010310}~(2023).
	
	\bibitem{haah_2022}
	Jeongwan Haah and Matthew~B. Hastings.
	\newblock ``Boundaries for the honeycomb code''.
	\newblock \href{https://dx.doi.org/10.22331/q-2022-04-21-693}{Quantum {\bf 6}, 693}~(2022).
	
	\bibitem{Gidney2022}
	Craig Gidney, Michael Newman, and Matt McEwen.
	\newblock ``Benchmarking the planar honeycomb code''.
	\newblock \href{https://dx.doi.org/10.22331/q-2022-09-21-813}{Quantum {\bf 6}, 813}~(2022).
	
	\bibitem{Gidney2022b}
	Craig Gidney.
	\newblock ``A pair measurement surface code on pentagons''.
	\newblock \href{https://dx.doi.org/10.22331/q-2023-10-25-1156}{Quantum {\bf 7}, 1156}~(2023).
	
	\bibitem{Kesselring2022}
	Markus~S. Kesselring, Julio~C. Magdalena de~la Fuente, Felix Thomsen, Jens
	Eisert, Stephen~D. Bartlett, and Benjamin~J. Brown.
	\newblock ``Anyon condensation and the color code''.
	\newblock \href{https://dx.doi.org/10.1103/PRXQuantum.5.010342}{PRX Quantum {\bf 5}, 010342}~(2024).
	
	\bibitem{Aguado2020}
	Ramón Aguado and Leo~P. Kouwenhoven.
	\newblock ``Majorana qubits for topological quantum computing''.
	\newblock \href{https://dx.doi.org/10.1063/PT.3.4499}{Physics Today {\bf 73}, 44}~(2020).
	
	\bibitem{Bombin2021}
	H\'ector Bomb\'{\i}n, Chris Dawson, Ryan~V. Mishmash, Naomi Nickerson, Fernando
	Pastawski, and Sam Roberts.
	\newblock ``Logical blocks for fault-tolerant topological quantum
	computation''.
	\newblock \href{https://dx.doi.org/10.1103/PRXQuantum.4.020303}{PRX Quantum	{\bf 4}, 020303}~(2023).
	
	\bibitem{gottesman2022opportunities}
	Daniel Gottesman.
	\newblock ``Opportunities and challenges in fault-tolerant quantum
	computation''
	\newblock \href{https://doi.org/10.48550/arXiv.2210.15844}{arXiv:2210.15844}~(2022).
	
	\bibitem{Coecke2011}
	Bob Coecke and Ross Duncan.
	\newblock ``Interacting quantum observables: categorical algebra and
	diagrammatics''.
	\newblock \href{https://dx.doi.org/10.1088/1367-2630/13/4/043016}{New Journal of Physics {\bf 13}, 043016}~(2011).
	
	\bibitem{Backens2014}
	Miriam Backens.
	\newblock ``The {ZX}-calculus is complete for stabilizer quantum mechanics''.
	\newblock \href{https://dx.doi.org/10.1088/1367-2630/16/9/093021}{New Journal of Physics {\bf 16}, 093021}~(2014).
	
	\bibitem{Coecke2017}
	Bob Coecke and Aleks Kissinger.
	\newblock ``Picturing quantum processes: A first course in quantum theory and
	diagrammatic reasoning''.
	\newblock \href{https://dx.doi.org/10.1017/9781316219317}{Cambridge University	Press}. ~(2017).
	
	\bibitem{Beaudrap2020}
	Niel de~Beaudrap and Dominic Horsman.
	\newblock ``The {ZX} calculus is a language for surface code lattice surgery''.
	\newblock \href{https://dx.doi.org/10.22331/q-2020-01-09-218}{Quantum {\bf 4}, 218}~(2020).
	
	\bibitem{Hadzihasanovic2018}
	Amar Hadzihasanovic, Kang~Feng Ng, and Quanlong Wang.
	\newblock ``Two complete axiomatisations of pure-state qubit quantum
	computing''.
	\newblock \href{https://dx.doi.org/10.1145/3209108.3209128}{Proceedings -
		Symposium on Logic in Computer SciencePages 502--511}~(2018).
	
	\bibitem{Jeandel2018}
	Emmanuel Jeandel, Simon Perdrix, and Renaud Vilmart.
	\newblock ``A complete axiomatisation of the {ZX}-calculus for {C}lifford+{T}
	quantum mechanics''.
	\newblock \href{https://dx.doi.org/10.1145/3209108.3209131}{Proceedings -
		Symposium on Logic in Computer SciencePages 559--568}~(2018).
	
	\bibitem{Wetering2020}
	John van~de Wetering.
	\newblock ``{ZX}-calculus for the working quantum computer scientist''
	\newblock \href{https://doi.org/10.48550/arXiv.2012.13966}{arXiv:2012.13966}~(2020).
	
	\bibitem{Peham_2022}
	Tom Peham, Lukas Burgholzer, and Robert Wille.
	\newblock ``Equivalence checking of quantum circuits with the {ZX}-calculus''.
	\newblock \href{https://dx.doi.org/10.1109/jetcas.2022.3202204}{{IEEE} Journal	on Emerging and Selected Topics in Circuits and SystemsPages 1--1}~(2022).
	
	\bibitem{Gidney2023}
	Matt McEwen, Dave Bacon, and Craig Gidney.
	\newblock ``Relaxing {H}ardware {R}equirements for {S}urface {C}ode {C}ircuits
	using {T}ime-dynamics''.
	\newblock \href{https://dx.doi.org/10.22331/q-2023-11-07-1172}{{Quantum} {\bf	7}, 1172}~(2023).
	
	\bibitem{Davydova2023}
	Margarita Davydova, Nathanan Tantivasadakarn, and Shankar Balasubramanian.
	\newblock ``Floquet codes without parent subsystem codes''.
	\newblock \href{https://dx.doi.org/10.1103/PRXQuantum.4.020341}{PRX Quantum	{\bf 4}, 020341}~(2023).
	
	\bibitem{Raussendorf2005}
	Robert Raussendorf, Sergey Bravyi, and Jim Harrington.
	\newblock ``Long-range quantum entanglement in noisy cluster states''.
	\newblock \href{https://dx.doi.org/10.1103/PhysRevA.71.062313}{Phys. Rev. A	{\bf 71}, 062313}~(2005).
	
	\bibitem{Nickerson2018}
	Naomi Nickerson and Hector Bombin.
	\newblock ``Measurement based fault tolerance beyond foliation''~(2018).
	
	\bibitem{Bombin2021a}
	Hector Bombin, Isaac~H. Kim, Daniel Litinski, Naomi Nickerson, Mihir Pant,
	Fernando Pastawski, Sam Roberts, and Terry Rudolph.
	\newblock ``Interleaving: Modular architectures for fault-tolerant photonic
	quantum computing''~(2021).
	
	\bibitem{Newman2020}
	Michael Newman, Leonardo~Andreta de~Castro, and Kenneth~R. Brown.
	\newblock ``Generating {F}ault-{T}olerant {C}luster {S}tates from {C}rystal
	{S}tructures''.
	\newblock \href{https://dx.doi.org/10.22331/q-2020-07-13-295}{{Quantum} {\bf 4}, 295}~(2020).
	
	\bibitem{Bombin2006}
	H.~Bombin and M.~A. Martin-Delgado.
	\newblock ``Topological quantum distillation''.
	\newblock \href{https://dx.doi.org/10.1103/PhysRevLett.97.180501}{Phys. Rev. Lett. {\bf 97}, 180501}~(2006).
	
	\bibitem{Bombin2018}
	Hector Bombin.
	\newblock ``2{D} quantum computation with 3{D} topological codes''
	\newblock \href{https://doi.org/10.48550/arXiv.1810.09571}{arXiv:1810.09571}~(2018).
	
	\bibitem{Breuckmann2021}
	Nikolas~P. Breuckmann and Jens~Niklas Eberhardt.
	\newblock ``Quantum low-density parity-check codes''.
	\newblock \href{https://dx.doi.org/10.1103/PRXQuantum.2.040101}{PRX Quantum	{\bf 2}, 040101}~(2021).
	
	\bibitem{Khesin2023}
	Andrey~Boris Khesin, Jonathan~Z Lu, and Peter~W Shor.
	\newblock ``Graphical quantum {C}lifford-encoder compilers from the {ZX} calculus''
	\newblock \href{https://doi.org/10.48550/arXiv.2301.02356}{arXiv:2301.02356}~(2023).
	
	\bibitem{Bombin2023}
	H{\'e}ctor Bomb{\'\i}n, Chris Dawson, Ye-Hua Liu, Naomi Nickerson, Fernando Pastawski, and Sam Roberts.
	\newblock ``Modular decoding: parallelizable real-time decoding for quantum	computers''
	\newblock \href{https://doi.org/10.48550/arXiv.2303.04846}{arXiv:2303.04846}~(2023).
	
	\bibitem{Litinski2022}
	Daniel Litinski and Naomi Nickerson.
	\newblock ``Active volume: An architecture for efficient fault-tolerant quantum computers with limited non-local connections''
	\newblock \href{https://doi.org/10.48550/arXiv.2211.15465}{arXiv:2211.15465} ~(2022).
\end{thebibliography}

\end{document}